\documentclass[aps,prl,floatfix,superscriptaddress,twocolumn,footinbib]{revtex4-1}
\usepackage{amssymb}
\usepackage{amsmath}
\usepackage{amsfonts}
\usepackage{appendix}
\usepackage{bm}
\usepackage{graphicx}
\usepackage{epsfig}
\usepackage{epstopdf}
\usepackage{balance}
\usepackage[dvipsnames]{xcolor}
\usepackage{calc}
\usepackage{natbib}
\usepackage[colorlinks,
            linkcolor=blue,
            anchorcolor=blue,
            citecolor=blue,
            urlcolor=blue]{hyperref}
\usepackage{lipsum}

\begin{document}
\title{Moir\'{e} Flat Bands of Twisted  Few-layer Graphite}

\affiliation{School of Physics and Wuhan National High Magnetic Field Center,
Huazhong University of Science and Technology, Wuhan 430074,  China}
\affiliation{International Center for Quantum Materials, School of Physics, Peking University, Beijing 100871, China}
\affiliation{Beijing Academy of Quantum Information Sciences, Beijing 100193, China}
\affiliation{CAS Center for Excellence in Topological Quantum Computation, University of Chinese Academy of Sciences, Beijing 100190, China}

\author{Zhen Ma}
\affiliation{School of Physics and Wuhan National High Magnetic Field Center,
Huazhong University of Science and Technology, Wuhan 430074,  China}
\author{Shuai Li}
\affiliation{School of Physics and Wuhan National High Magnetic Field Center,
Huazhong University of Science and Technology, Wuhan 430074,  China}
\author{Meng-Meng Xiao}
\affiliation{School of Physics and Wuhan National High Magnetic Field Center,
Huazhong University of Science and Technology, Wuhan 430074,  China}
\author{Ya-Wen Zheng}
\affiliation{School of Physics and Wuhan National High Magnetic Field Center,
Huazhong University of Science and Technology, Wuhan 430074,  China}
\author{Ming Lu}
\affiliation{Beijing Academy of Quantum Information Sciences, Beijing 100193, China}
\affiliation{International Center for Quantum Materials, School of Physics, Peking University, Beijing 100871, China}
\author{HaiWen Liu}
\affiliation{Center for Advanced Quantum Studies, Department of Physics, Beijing Normal University, Beijing 100875, China}
\author{Jin-Hua Gao}
\email{jinhua@hust.edu.cn}
\affiliation{School of Physics and Wuhan National High Magnetic Field Center,
Huazhong University of Science and Technology, Wuhan 430074,  China}
\author{X. C. Xie}
\affiliation{International Center for Quantum Materials, School of Physics, Peking University, Beijing 100871, China}
\affiliation{Beijing Academy of Quantum Information Sciences, Beijing 100193, China}
\affiliation{CAS Center for Excellence in Topological Quantum Computation, University of Chinese Academy of Sciences, Beijing 100190, China}
\begin{abstract}
We report that the twisted few layer graphite (tFL-graphite) is a new family of moir\'{e}  heterostructures (MHSs), which has richer and highly tunable moir\'{e} flat band structures entirely distinct from all the known MHSs.  A tFL-graphite is composed of two few-layer graphite (Bernal stacked multilayer graphene), which are stacked on each other with a small twisted angle. The moir\'{e} band structure of the tFL-graphite strongly depends on the layer number of its composed two van der Waals layers.  Near the magic angle, a tFL-graphite always has two nearly flat bands coexisting with a few pairs of narrowed dispersive (parabolic or linear) bands at the Fermi level, thus, enhances the DOS at $E_F$. This coexistence property may also enhance the possible superconductivity as been demonstrated in other multiband superconductivity systems. Therefore, we expect strong multiband correlation effects in tFL-graphite. Meanwhile,  a proper perpendicular electric field can induce several isolated nearly flat bands with nonzero valley Chern number in some simple tFL-graphites, indicating that tFL-graphite is also a novel topological flat band system.
\end{abstract}
\maketitle
\emph{Introduction.}---Moir\'{e} heterostructures~(MHSs) have drawn great research interest  recently~\cite{cao2018a,cao2018b,Yankowitz2019,Lu2019653,Sharpe2019605,lau2019,Jiang201991,wyao2016}. The celebrated example is the twisted bilayer graphene (TBG), in which two atomically thin van der Waals (vdW) layers, \textit{i.e.}~graphene monolayers here, are stacked with a controlled twist angle $\theta$. A small twist angle results in a  long period moir\'{e} superlattice. Most importantly, the moir\'{e} interlayer hopping (MIH) will give rise to two nearly flat bands, when $\theta$ approaches the so-called magic angles~\cite{prl2007,mac2011,prb2012,Koshino2013,koshino2018,kangjian2018}.  Due to the divergent DOS of the moir\'{e} flat bands,  exotic correlation phenomena,  \textit{e}.\textit{g}.~superconductivity and Mott insulator, have been observed in experiments~\cite{cao2018b,cao2018a,Yankowitz2019,Lu2019653,Sharpe2019605,lau2019,Jiang201991}, and intensively studied with various theoretical methods~\cite{lado2017,xu2018,senthilprx,fuprx,Phillips,Guinea2018,yangfan2018,guohuaiming,rahul,wusuperconductor,lianbiao,meanfield2018,strongSC,You2019}.

Further studies indicate that the moir\'{e} flat bands  exist in a variety of MHSs, such as  twisted double bilayer graphene~\cite{2019arXiv190308130L,zhangguangyu2019,cao2019,tutu2019,yhzhang2019,Koshino,jeil2019,lee2019,wu2019,Wudoublebilayer,DFTdoublebilayer}, twisted trilayer graphene~\cite{jinhuaGao2019,helin2018,brey2013,lixiaotrilayer,kaxiras2019,shi2020,chenshaowen2020,young2020}, twisted   rhombohedral (ABC) multilayer  graphene~\cite{liu20192}, trilayer graphene on boron nitride~\cite{jeilprl2019,wangfeng2019,Chennature} and transition metal dichalcogenide heterostructures~\cite{wutmd2018,tmd2018,panyi,wufengcheng2019,tmdprb2019,sunjinhua2019}. Note that, in all the above-mentioned MHSs, the low energy moir\'{e} band structures are in some sense similar, \textit{i}.\textit{e}.,  two  moir\'{e} flat bands at the charge neutrality point isolated from other high energy bands.
 This is because that their vdW layers, like graphene monolayer, bilayer, ABC stacked multilayer, \textit{etc}., just happen to have two low energy bands near the Fermi level $E_F$, which are hybridized  by the MIH to form the two flat bands. In this work, we report  a new family of moir\'{e} flat band systems, \emph{twisted few-layer graphite} (tFL-graphite), which have richer and highly controllable moir\'{e} flat band structures entirely distinct from all the known MHSs.


\begin{figure}[t!]
\centering
\includegraphics[width=8.5cm]{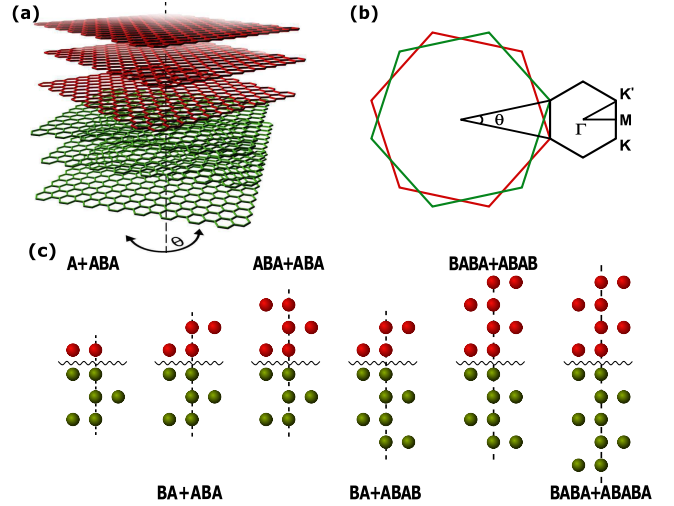}
\caption{(a) Schematic of the tFL-graphite, red (green) is the top (bottom) vdW layer. (b) Moir\'{e} Brillouin zone (black), red (green) is the first BZ of the top (bottom) vdW layer. (c) Side view of several tFL-graphite configurations.}
\label{fig1}
\end{figure}

\begin{figure*}[t!]
\centering
\includegraphics[width=17.5cm]{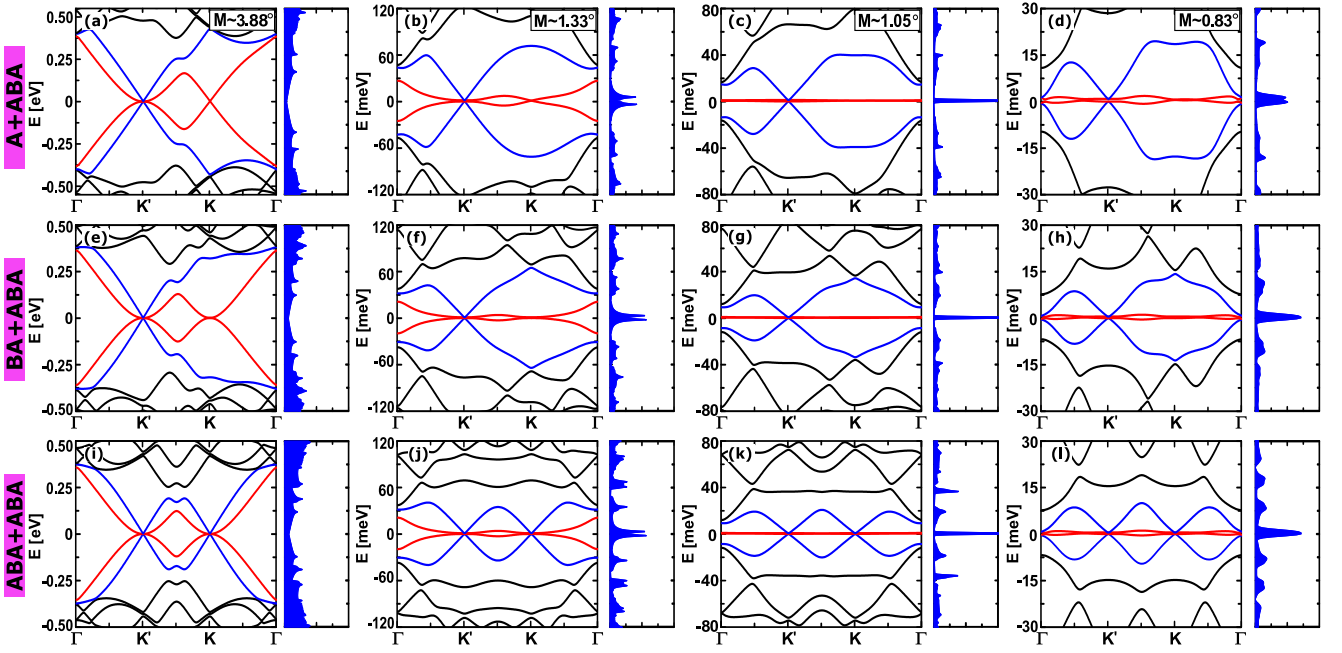}
\caption{(a-d), (e-h) and (i-l) are the moir\'{e} bands of the A+ABA, BA+ABA and ABA+ABA, respectively. (a, e, i) are calculated at $\theta=3.88^\circ$; (b, f, j) are calculated at $\theta=1.33^\circ$; (c, g, k) are calculated at $\theta=1.05^\circ$; (d, h, l) are calculated at $\theta=0.83^\circ$. The moir\'{e} bands of AB+ABA and BAB+ABA are given in Ref.~\onlinecite{Supplement}.  Parameters:  $\omega_1/\omega_2/\gamma_0/\gamma_1/\gamma_3/\gamma_4=78/98/2610/360/0/0\ meV$, $\xi=-1$~\cite{liu20192,lee2019}.}
\label{fig2}
\end{figure*}

Typical tFL-graphites are shown in Fig.~\ref{fig1}.
Here,  few layer graphite (FL-graphite) refers to  multilayer graphene with  Bernal (AB) stacking, in contrast to that with rhombohedral one. Generally, a M+N tFL-graphite is composed of two FL-graphites (red and green in Fig.~\ref{fig1}) stacked on top of each other with a twist angle $\theta$, where M (N) is the layer number of the top (bottom) FL-graphite.  We assume that N$\geq$M without loss of generality, and only focus on the cases with  N$\ge$3.
So, the simplest tFL-graphite here is the A+ABA (or 1+3) configuration, where A is the top monolayer (red) and ABA is the bottom ABA stacked trilayer (green) [see Fig.~\ref{fig1} (c)].

The distinct moir\'{e} band structure of tFL-graphite mainly results from the richer electronic structure of FL-graphite.  It is  known that a N-layer FL-graphite has $N/2$ electron-like bands and $N/2$ hole-like bands touching at the Dirac points if $N$ is even, while an additional pair of linear bands appears for an odd N~\cite{guinea2006,koshino2007,minhongki2008}.
Thus, it is natural to expect a rather different moir\'{e} band structure in the tFL-graphite.
And, as we will see later, twist actually has different influences on the bands of FL-graphite.

We theoretically calculate the electronic structures of tFL-graphites based on a continuum model.  Our results show that the electronic structure of a tFL-graphite  strongly depends on  the layer number of its composed vdW layers,  which always has two moir\'{e} flat bands coexisting with a few pair of parabolic (or linear) narrow bands near the magic angle. For  a M+N tFL-graphite, the main characteristics of its moir\'{e} band structures  are summarized as follows: (1) The number of the moir\'{e} bands at $E_F$ is determined by N, \textit{i.e.}~the layer number of the thicker vdW layer, where it has N (N+1) moir\'{e} bands if N is even (odd); (2) Near the magic angle (about $1.05^\circ$), two of these moir\'{e} bands become  flat, while others are still parabolic or linear at the Dirac points with narrowed bandwidth; (3) We can get four isolated nearly flat bands with various nonzero valley Chern number by applying a proper electric field in the N=3 cases.

Due to the unique electronic structures, the tFL-graphite has several great advantages as an intriguing MHS. (1) It has a richer and more flexible moir\'{e} band structure than the known MHSs, which can be dramatically changed by choosing different layer number. (2) The coexistence of flat bands and other dispersive bands  will not only enhance the density of states (DOS) at $E_F$, but also may meet the demand for the enhancement of superconductivity according to the steep band/flat band scenario~\cite{Simon1997,antonio2019}. Both these two points imply that stronger multiband correlation effects, \textit{e}.\textit{g}. superconductivity at higher temperature, may occur in the tFL-graphite. (3) It is a promising platform to study correlation effects in topological bands.
(4) The sample preparation of tFL-graphite may not be too challenging~\cite{tan2014}. Note that natural graphite is  Bernal stacked (more stable than the ABC graphene multilayers), and FL-graphite can be directly obtained by mechanical exfoliation.

\emph{Continuum model.}---
Similar as the TBG,  the required $\theta$ in the commensurate cases is determined by an integer $m$, where $\cos \theta=(3m^2+3m+\frac{1}{2})/(3m^2+3m+1)$.
The corresponding lattice vectors of the moir\'{e} supercell are $\bm{t}_1=m\bm{a}_1+(m+1)\bm{a}_2$ and $\bm{t}_2=-(m+1)\bm{a}_1+(2m+1)\bm{a}_2$, while $\bm{a}_1=a(1/2,\sqrt{3}/2)$ and $\bm{a}_2=a(-1/2,\sqrt{3}/2)$ are the lattice vectors of graphene with lattice constant $a\approx0.246\ nm$. The moir\'{e} Brillouin zone (BZ) is given in Fig.~\ref{fig1} (b) (black), which is determined by the BZ of the top vdW layer (red) and that of the bottom layer (green). The reciprocal lattice vectors $G_i$ can be obtained from $\bm{G}_i\cdot \bm{t}_j=2\pi\delta_{ij}$, and $K$ ($K'$) is the Dirac point corresponding to the top (bottom) layer [see Fig.~\ref{fig1} (b)].

The effective continuum model is used to describe the tFL-graphite~\cite{mac2011,prb2012,Koshino2013}. For a M+N tFL-graphite, the Hamiltonian is
\begin{equation}\label{eq1}
H_{M+N}(\theta)=\left( \begin{array}{cc}
H_{N}(k_1)& T(r)   \\
T^\dagger(r)&H_{M}(k_2)\\
\end{array} \right)+U
\end{equation}
where $k_1=R(-\theta/2)(k-K)$, $k_2=R(\theta/2)(k-K')$ and $R(\theta)$ is the rotation matrix. $H_M$ ($H_N$) is the Hamiltonian of M-layer (N-layer) FL-graphite~\cite{guinea2006,koshino2007,minhongki2008,zhangfan2010}.  For example, $H_{N=3}$ is the Hamiltonian of the ABA trilayer
\begin{equation}\label{H1}
H_{N=3}(k)=\left( \begin{array}{ccc}
h_0(k) & g(k)^\dagger&0   \\
 g(k)& h_0(k)&g(k)\\
 0&g(k)^\dagger&h_0(k)
\end{array} \right),
\end{equation}
where $h_0(\bm{k})=-\hbar v_F\bm{k}\cdot\bm{\sigma}$ is the low-energy effective Hamiltonian for graphene monolayer and  $g(k)$ is the interlayer hopping
\begin{equation}\label{gk}
g(k)=\left( \begin{array}{cc}
\hbar v_4k_+ &\gamma_1\\
\hbar v_3k_- &\hbar v_4k_+
\end{array} \right).
\end{equation}
Here, $k_{\pm}=\xi k_x\pm ik_y$, with $\xi=\pm 1$ for the two different valley in multilayer graphene. $\gamma_1$ is the vertical hopping and $v_i=\frac{\sqrt{3}\gamma_i a}{2\hbar}$ (i=3,4).  $\gamma_3$ and $\gamma_4$ are the remote interlayer hoppings, which stand for the trigonal warping and electron-hole asymmetry, respectively.
The moir\'e interlayer coupling is $T(r)=\sum_{n=0,1,2}T_n\cdot e^{-iQ_n\cdot r}$, where
\begin{equation}\label{H1}
T_{n}= I_{MN}\otimes\left( \begin{array}{ccc}
\omega_1&\omega_2e^{in\phi}\\
\omega_2e^{-in\phi}&\omega_1
\end{array} \right).
\end{equation}
Here, $I_{MN}$ is a $N \times M$ matrix with only one nonzero matrix element $I_{MN}(1,N)=1$, $\phi=2\pi/3$, $Q_n=R(n\phi)\cdot(K-K')$. At last, $U$  in Eq.~\eqref{eq1} is a diagonal matrix denoting the perpendicular electric field induced potential in layers, where we set the potential difference between adjacent two layers is $V$~\cite{Supplement}.

\emph{Moir\'{e} bands of the simplest tFL-graphite.}---Most of the moir\'{e} band characteristics of tFL-graphite can be seen from the simplest tFL-graphites, \textit{i}.\textit{e}., N=3 and M=(1,2,3). The calculated moir\'{e} bands are given in Fig.~\ref{fig2},  each row of which corresponds to one structure of tFL-graphite. 
Here, we use a minimal model, only including the dominating interlayer hopping $\gamma_1$ in Eq.~\eqref{gk}, to illustrate the ideal flat bands at the magic angle. We also use a full parameter model  to account for the realistic situations, where the influence of $\gamma_3$ and $\gamma_4$ are considered~\cite{Supplement}.
The first to fourth columns in Fig.~\ref{fig2} are the moir\'{e} bands with $\theta=3.88^\circ$,  $\theta=1.33^\circ$, $\theta=1.05^\circ$ and $\theta=0.83^\circ$, respectively. Corresponding DOS is also plotted.

\begin{figure}[t!]
\centering
\includegraphics[width=8.5cm]{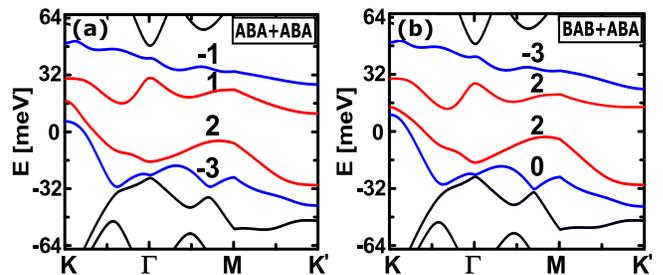}
\caption{(a), (b) are the moir\'{e} bands under a perpendicular electric field for the ABA+ABA and BAB+ABA, respectively. $V=20$ meV, $\theta=1.33^{\circ}$ and full parameter model is used. The black number are the valley Chern numbers for four moir\'{e} bands (red and blue solid lines).}
\label{fig3}
\end{figure}

\begin{figure*}[t!]
\centering
\includegraphics[width=17.5cm]{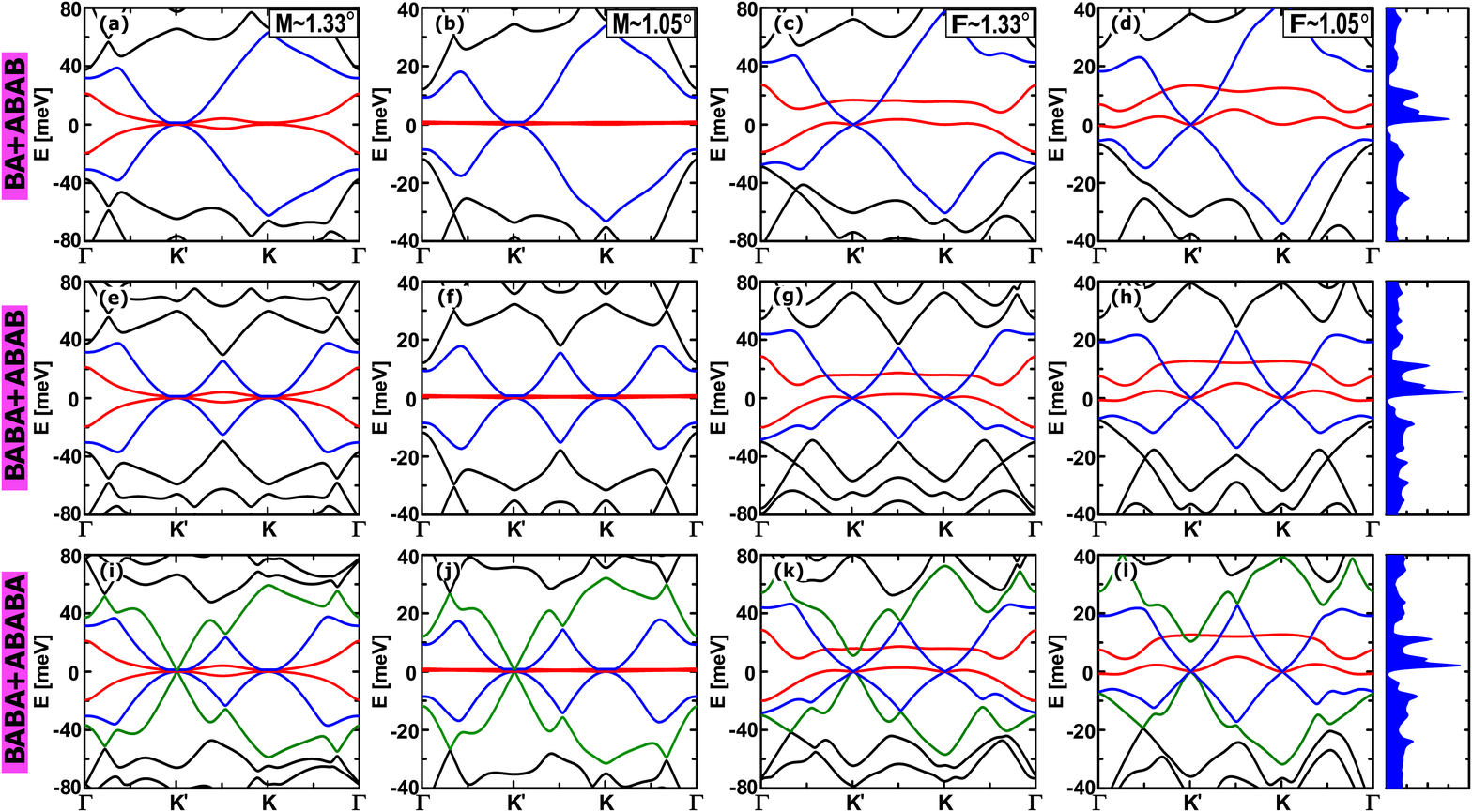}
\caption{(a-d), (e-h) and (i-l) are the moir\'{e} bands of the BA+ABAB, BABA+ABAB and BABA+ABABA, respectively. (a, e, i) are calculated with the minimal model at $\theta=1.33^\circ$; (b, f, j) are calculated with minimal model at $\theta=1.05^\circ$; (c, g, k) are calculated with full parameter model at $\theta=1.33^\circ$; (d, h, l) are calculated with full parameter model at $\theta=1.05^\circ$, and the corresponding DOS is also plotted. Parameters: $\omega_1/\omega_2/\gamma_0/\gamma_1/\gamma_3/\gamma_4=78/98/2610/360/283/138\ meV$, $\xi=-1$.}
\label{fig4}
\end{figure*}

First of all,  tFL-graphite has more moir\'{e} bands than all the known MHSs at $E_F$. For example, all the cases (N=3) in Fig.~\ref{fig2} have four moir\'{e} bands at $E_F$, instead of two like in TBG. It is because that the low energy moir\'{e} bands of a tFL-grapite are constructed by hybridizing the bands of top and bottom vdW layers near the Dirac points via the MIH. In TBG,  each vdW layer has two linear bands near the Dirac point.  But in the A+ABA tFL-graphite,  the top vdW layer (monolayer) has two linear bands, while the bottom vdW layer (ABA-trilayer) has  four bands,  \textit{i}.\textit{e}.~a  parabolic electron band,  a parabolic hole band and  a pair of linear bands.    As shown in Fig.~\ref{fig2} (a-d), the MIH hybridizes the two parabolic bands of the  bottom ABA-trilayer with the two linear bands of the top monolayer (red lines), while the remaining  two linear bands of the ABA-trilayer are modified into two other moir\'{e} bands at $E_F$ (blue lines).  Thus, there are four moir\'{e} bands near $E_F$ in the A+ABA tFL-graphite.
The moir\'{e} bands of BA+ABA [Fig.~\ref{fig2} (e-h)] and ABA+ABA [Fig.~\ref{fig2} (i-l)] tFL-graphites can be understood in a similar way.  Generally, for a M+N tFL-graphite, the number of moir\'{e} bands at $E_F$ is equal to that of the thicker vdW layer, \textit{i}.\textit{e}. it has $N$  moir\'{e} bands ($N+1$) if N is even (odd) when $N \geq M$.  Half of the moir\'{e} bands are electron-like and the others are hole-like.

The magic angle of the tFL-graphite is  about $1.05^\circ$, the same as that of TBG. However, only two of these mori\'{e} bands in tFL-graphite can be transformed into flat bands at the magnic angle,  while others are still dispersive but their bandwidth  are greatly narrowed. This is shown clearly in Fig.~\ref{fig2} (c), (g),  (k), which are calculated with $\theta=1.05^\circ$  based on the minimal model. Increasing $\theta$, the bandwidth of the moir\'{e} bands becomes larger, and the the flat bands become dispersive [see Fig.~\ref{fig2} (b), (f), (j) with $\theta=1.33^\circ$].
With a twist angle  smaller than the magic angle [\textit{e.g.}~$\theta=0.83^\circ$ in Fig.~\ref{fig2} (d), (h), (l)],  the flat band  [red lines] will become slightly dispersive while the bandwidth of the other two moir\'{e} bands [blue lines] are reduced further [smaller than 10 meV in Fig.~\ref{fig2} (l)]. It suggests that the correlation in such narrow dispersive bands may be also very strong with small $\theta$.
At large twist angle [\textit{e.g.}~Fig.~\ref{fig2} (a), (e), (i) with $\theta=3.88^\circ$], the shape of the bands near $K$ ($K'$) points recovers to that of the isolated top (bottom) vdW layer.
Note that  the A+ABA [Fig.~\ref{fig2} (c)] and BA+ABA [Fig.~\ref{fig2} (g)] both host flat bands coexisting with a single Dirac cone ($K'$), while ABA+ABA [Fig.~\ref{fig2} (k)] have two Dirac cones.  This coexistence feature is very interesting because that  the simultaneous occurrence of flat and steep bands is a favorable condition to  enhance superconductivity, according to the ``steep band/flat band'' scenario of superconductivity~\cite{Simon1997,antonio2019}. Note that coexistence of flat band with linear bands at $E_F$ was also noticed in some multi-twist MHSs very recently\cite{lixiaotrilayer,kaxiras2019,multitwist2019}, while only one twist is required here.

The low energy moir\'{e} bands of tFL-graphite also exhibit exotic topological properties, the valley Chern number of which can be nonzero. In Fig.~\ref{fig3},  we consider two example, \textit{i.e.}~ABA+ABA and BAB+ABA. We first see that the four moir\'{e} bands at $E_F$ can be isolated by a proper perpendicular electric field, while their bandwidth are still small. For these isolated moir\'{e} bands, we calculate their valley Chern numbers with the standard formula~\cite{Supplement}, which are indicated by black numbers in Fig.~\ref{fig3}.  The first interesting issue is that tFL-graphite here have four moir\'{e} bands with nonzero valley Chern number, while other known MHSs only have two~\cite{yhzhang2019,jinhuaGao2019,liu20192, jeilprl2019,Koshino}.  Furthermore, despite that ABA+ABA and BAB+ABA differ only in the relative orientation and have very similar band structure~\cite{Supplement}, their moir\'{e} bands have  different valley Chern number. The cases of AB+ABA and BA+ABA are similar, and the valley Chern number is controllable by the twist angle and perpendicular electric field~\cite{Supplement}.

\emph{Moir\'{e} bands of thicker tFL-graphite.}---Now we study the moir\'{e} bands of thicker tFG-graphites. Several typical examples are given in Fig.~\ref{fig4}.
These thicker tFL-graphites have  various moir\'{e} bands near the $E_F$, depending on their own layer numbers. For example, with N=4 (BA+ABAB, BABA+ABAB), there exists a pair of flat bands (red lines) coexisting with a parabolic electron and a parabolic hole bands (blue lines), as shown in Fig.~\ref{fig4} (a-b) and (e-f).  Compared with the N=3 cases in Fig.~\ref{fig2}, the parabolic bands here can obviously enhance the DOS at $E_F$, which indicates that correlation effects in thicker tFL-graphite may be stronger. Increasing the layer number further, it is able to induce more moir\'{e} bands near the $E_F$.  A N=5 example is given in Fig.\ref{fig4} (i-j), which has six moir\'{e} bands including a pair of flat bands, two parabolic bands and two linear bands (green lines).

Note that the influence of the remote hopping $\gamma_3$ and $\gamma_4$ is critical, and  different  on various moir\'{e} bands, as illustrated by the full parameter model calculations in Fig.~\ref{fig4} (c-d), (g-h) and (k-l). The remote hopping always separates the  the flat bands in energy (red lines) and make them dispersive. But it does not significantly modify the parabolic bands near the Dirac points (blue lines).  Meanwhile, a gap at the Dirac points is induced to the linear bands in N=5 case [see in Fig.~\ref{fig4} (k-l), green lines].


\emph{Summary.}---Our numerical calculations reveal and give some intuitive understanding about the unique, richer and highly controllable the moir\'{e} flat band structure in tFL-graphite, which may be directly verified by the nano-ARPES technique~\cite{chen2017,arpeswang,thompson2020,lisi2020,jones2020}.  However, the research on the moir\'{e} bands in tFL-graphite is still at early stage,  though the FL-graphite has been studied for rather a long time~\cite{tan2014,multilayer2018,nano2019,naturecommu2015,science2018}.  Many basic problems are still unknown. For example, why does twist have different influences on the bands of FL-graphite? Whether can the multiband feature here  significantly enhance the superconductivity? We hope our study will stimulate further interest on this promising system with moir\'{e} multiband correlations.

\begin{acknowledgments}
We thank Jinhua Sun for helpful discussion. This work is supported  by the National Natural Science Foundation of China (Grants No.~11534001, 11874160, 11274129, 11874026, 61405067), and the Fundamental Research Funds for the Central Universities (HUST: 2017KFYXJJ027), and NBRPC (Grants No. 2015CB921102).
\end{acknowledgments}
\bibliography{twistedgraphene1}
\newpage
\clearpage
\onecolumngrid
\section*{SUPPLEMENTARY Materials OF ``Moir\'{e} Flat Bands of Twisted  Few-layer Graphite''}
\begin{appendix}
	\setcounter{equation}{0}
	\setcounter{figure}{0}
	\setcounter{table}{0}
	\setcounter{page}{1}
	\makeatletter
	\renewcommand{\theequation}{S\arabic{equation}}
	\renewcommand{\thefigure}{S\arabic{figure}}
	\renewcommand{\bibnumfmt}[1]{[S#1]}

\section{I. The term to describe perpendicular electric field in the Hamiltonian of tFL-graphite}
In Eq.~1 of the main text, we give the Hamiltonian of the tFL-graphite, where the term $U$ is a diagonal matrix to describe the perpendicular electric field.  The effect of a perpendicular electric field is to induce potential difference between layers. We ignore the screening effect and  assume that the potential difference distributes uniformly between the top and bottom layers. And, as mentioned in the main text, we set $V$ as the potential difference between two adjacent layers. Let us take the case of 2+4 tFL-graphite for example.
Since there are six layers in  2+4 tFL-graphite, the total potential difference between the top and bottom layers is thus 5V. We then set the potential of top (bottom) layer is 2.5V (-2.5V). So, U=diag(2.5V, 2.5V, 1.5V, 1.5V, 0.5V, 0.5V, -1.5V, -1.5V, -2.5V, -2.5V).
Note that there are two Bloch basis for each layer, \textit{i.e.}, one is from sublattice A and the other is from sublattice B. So, they have the same potential.

\section{II. Complementary results of tFL-graphite with N=3}

\begin{figure}[h!]
\centering
\includegraphics[width=5cm]{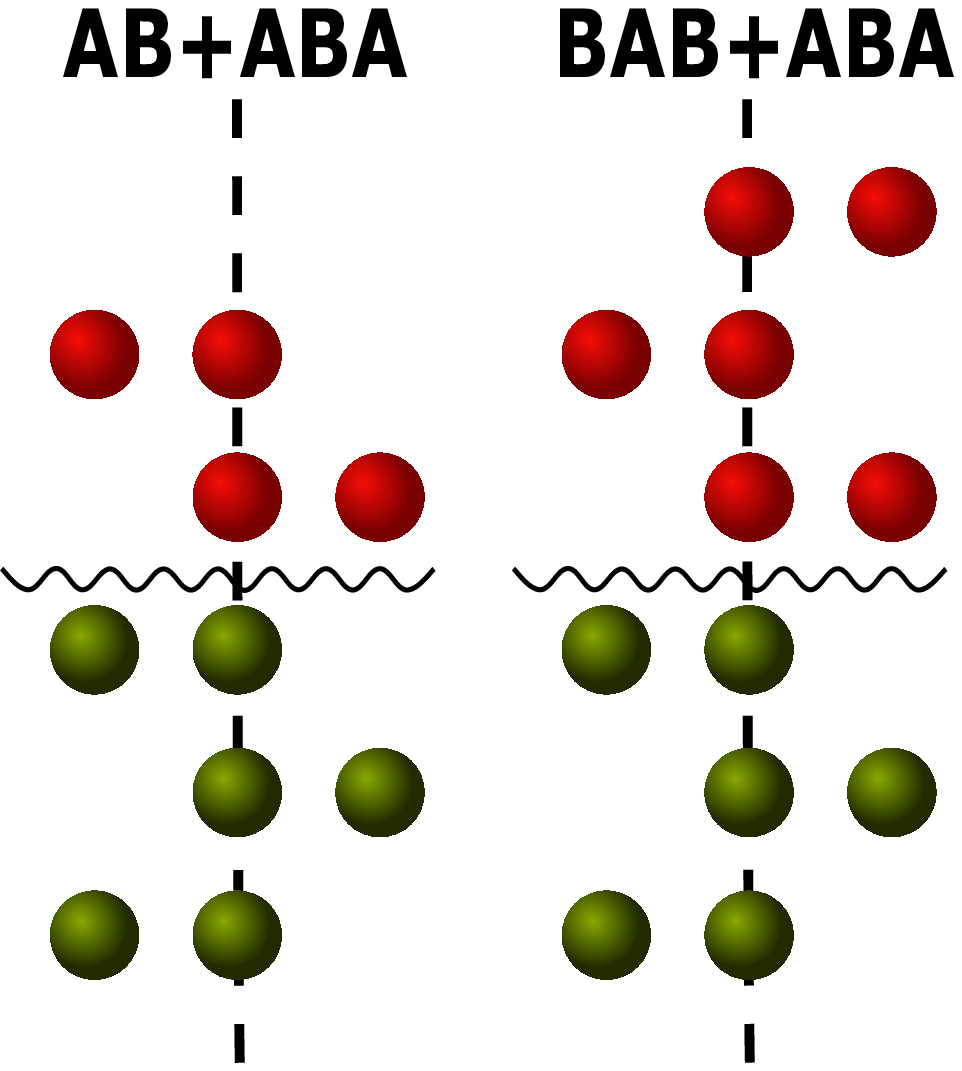}
\renewcommand\thefigure{S\arabic{figure}}
\caption{Side view of tFL-graphite configurations: AB+ABA, BAB+ABA.}
\label{figs1}
\end{figure}

\begin{figure}[h!]
\centering
\includegraphics[width=17cm]{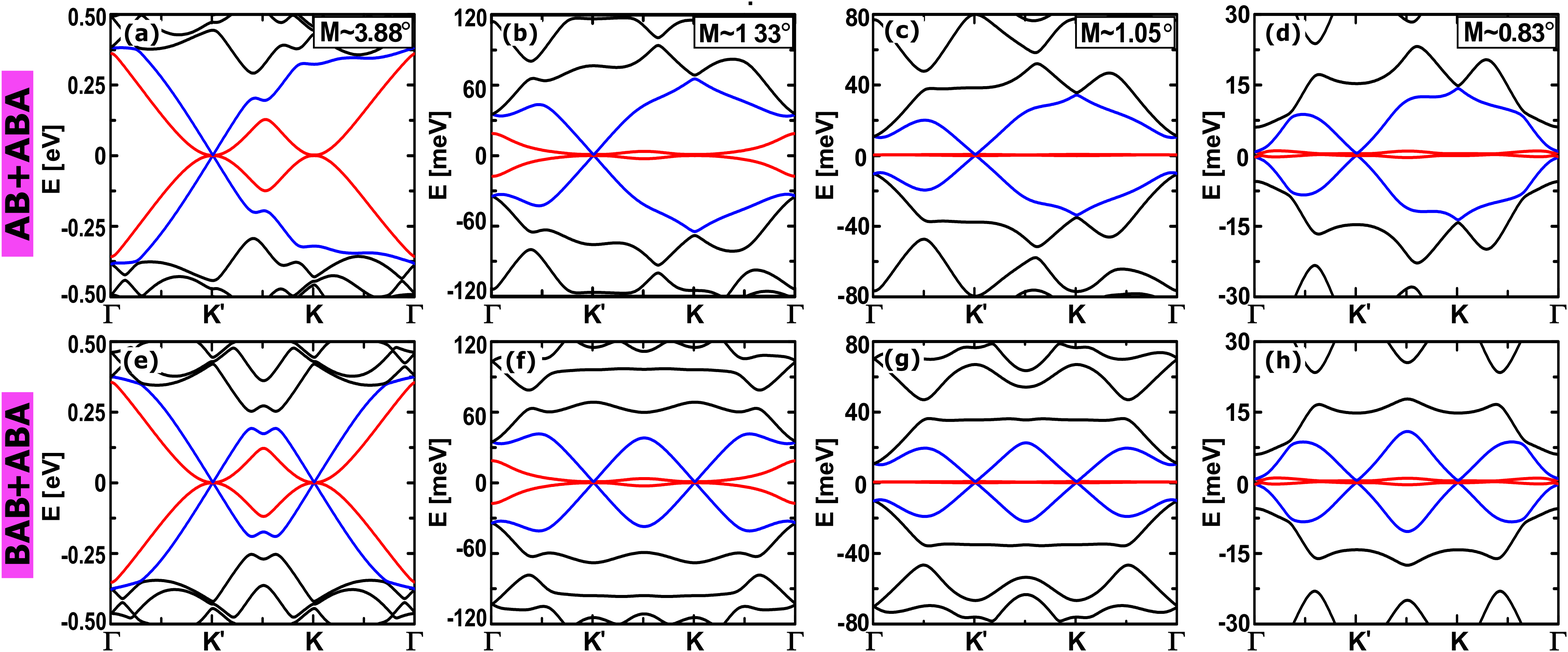}
\renewcommand\thefigure{S\arabic{figure}}
\caption{(a-d) and (e-h) are the moir\'e bands of the AB+ABA and BAB+ABA, respectively. (a,e) are calculatedat $\theta=3.88^\circ$. (b,f) are calculated at $\theta=1.33^\circ$. (c,g) are calculated at $\theta=1.05^\circ$. (d,h) are calculated at $\theta=0.83^\circ$. Minimal model is used and the parameters are the same as Fig.~2 of the main text.}
\label{figs2}
\end{figure}

\emph{Moir\'{e} bands of AB+ABA and BAB+ABA tFL-graphite}---In the main text, we show the moir\'e bands of the A+ABA, BA+ABA and ABA+ABA in the Fig.~2. Note that there are two remaining configurations of tFL-graphite with $N=3$, \textit{i.e.}, AB+ABA and BAB+ABA as shown in Fig.~\ref{figs1}. Their moir\'{e} bands are given in Fig.~\ref{figs2}.  The only difference between AB+ABA and BA+ABA is the relative orientation of the vdW layers, and their moir\'{e} band structures are  slightly different.  For example, for the AB+ABA [Fig.~\ref{figs2} (b)], two low energy bands (the blue line and the black line) are touching at $\Gamma$ point, while there is an observable gap in BA+ABA [Fig.~2 (f) of the main text]. The moir\'{e} bands of ABA+ABA and BAB+ABA configurations are also slightly different at the $\gamma$ point.

\vbox{}

\begin{figure}
\centering
\includegraphics[width=17.5cm]{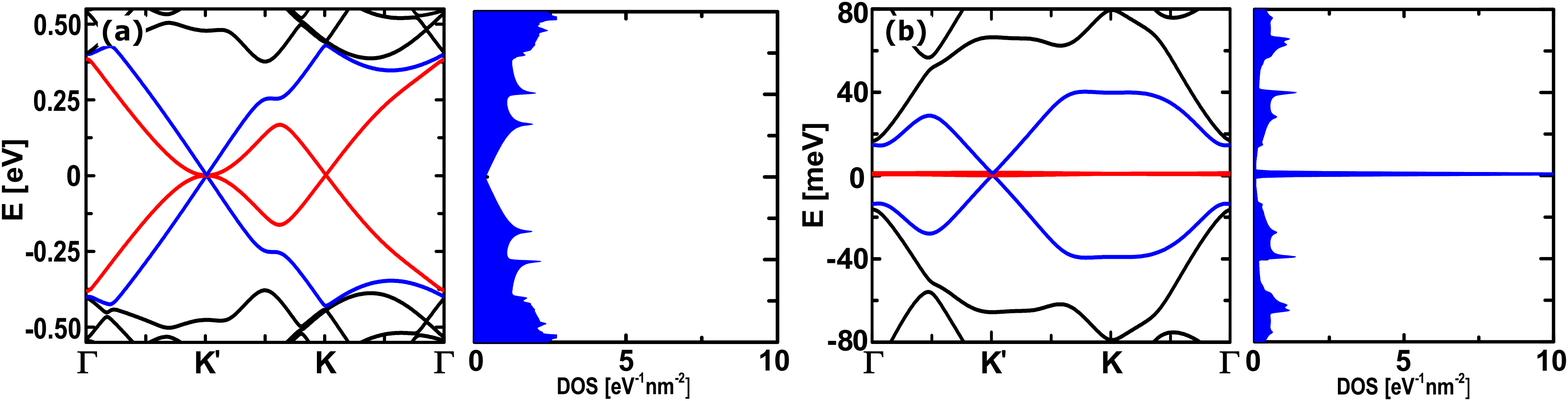}
\renewcommand\thefigure{S\arabic{figure}}
\caption{(a) and (b) are the moir\'e band (and DOS) of the A+ABA  at $\theta=3.88^\circ$ and $\theta=1.05^\circ$, respectively. Minimal model is used.}
\label{figs3}
\end{figure}

\emph{Unit of the DOS}---In order to give an intuitive impression about the order of magnitude of the DOS in the tFL-graphite [see Fig.~2 of the main text], we  replot the bands and DOS of A+ABA in Fig.~\ref{figs3}, where the unit of DOS is given.

\begin{figure}
\centering
\includegraphics[width=14cm]{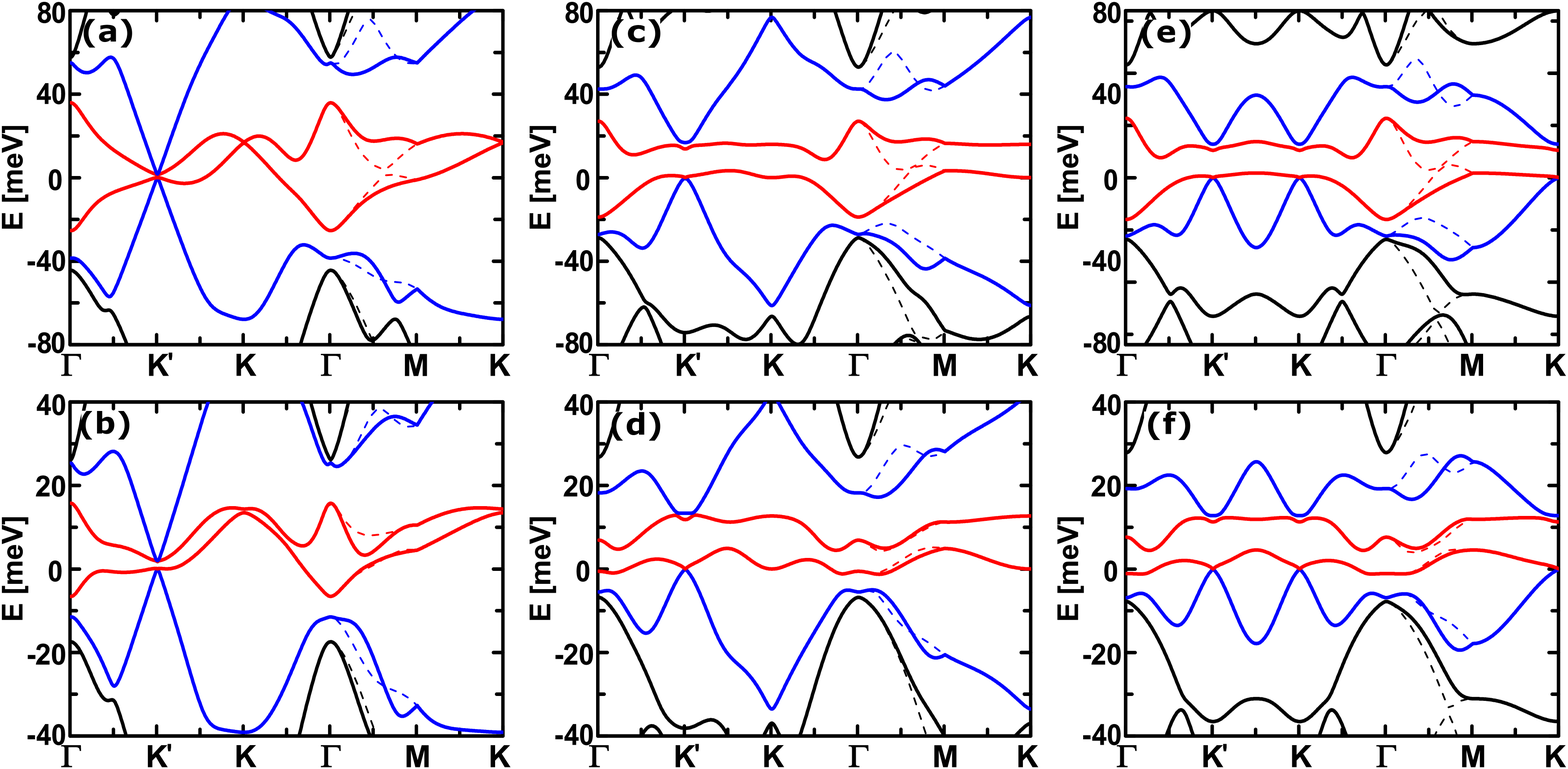}
\renewcommand\thefigure{S\arabic{figure}}
\caption{ Moir\'{e} bands calculated with full parameter model. (a-b) are for A+ABA, (c-d) are for AB+ABA and (e-f) are for ABA+ABA. (a), (c), (e) are calculated at $\theta=1.33^\circ$, and (b), (d), (f) are calculated at $\theta=1.05^\circ$.  $\gamma_0/\gamma_1/\gamma_3/\gamma_4=2610/360/283/138$ meV. }
\label{figs4}
\end{figure}

\begin{figure}[!h]
\centering
\includegraphics[width=14cm]{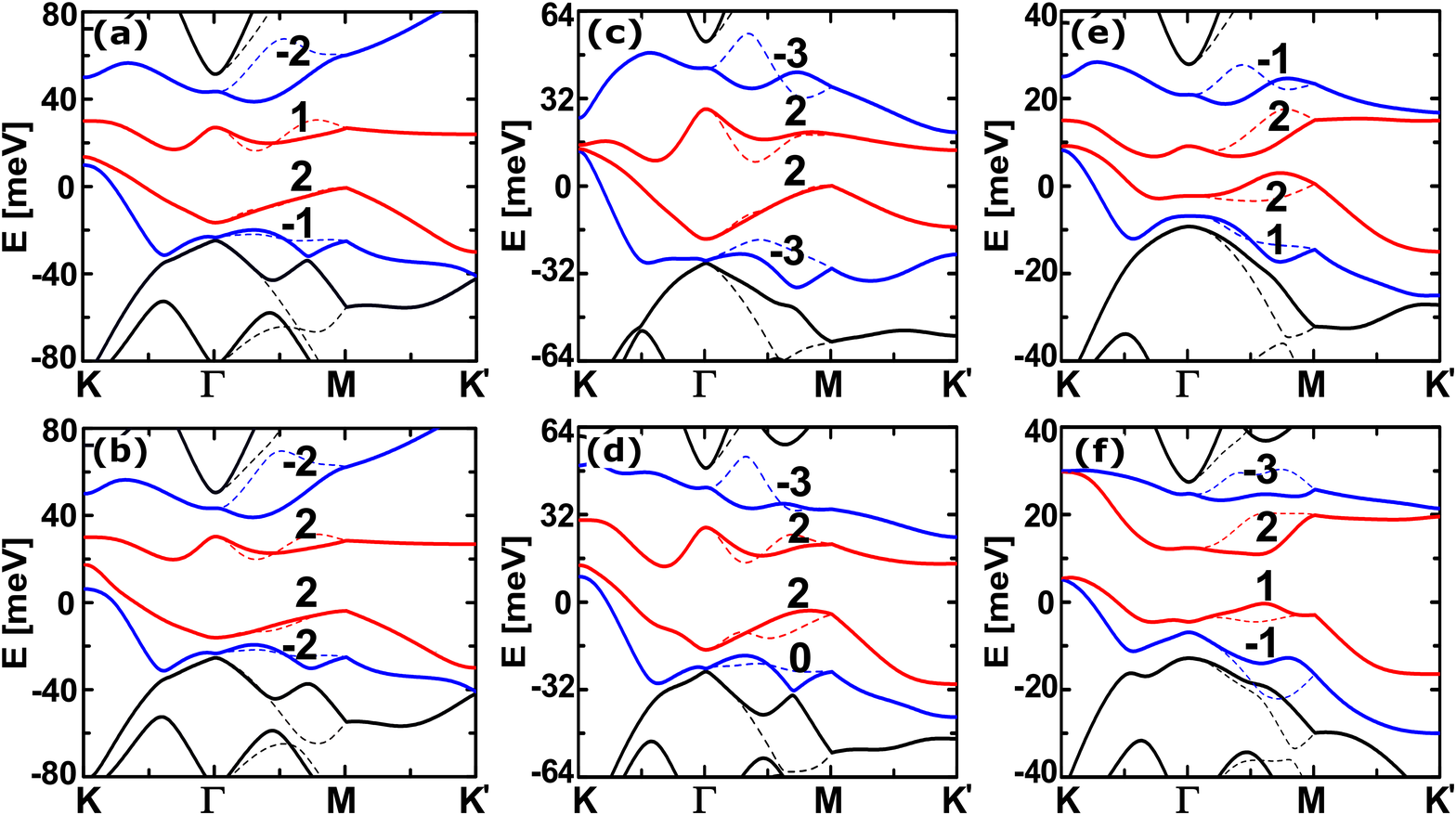}
\renewcommand\thefigure{S\arabic{figure}}
\caption{Moir\'{e} bands and valley Chern number in the presence of perpendicular electric field.  (a) AB+ABA with $V=20$ meV and $\theta=1.33^\circ$. (b) BA+ABA with $V=20$ meV and $\theta=1.33^\circ$. (c) ABA+ABA with $V=10$ meV and $\theta=1.33^\circ$. (d) ABA+ABA with $V=20$ meV and $\theta=1.33^\circ$. (e) ABA+ABA with $V=10$ meV and $\theta=1.05^\circ$. (f) ABA+ABA with $V=20$ meV and $\theta=1.05^\circ$. The full parameter model is used. Valley Chern numbers are denoted by black numbers and the dashed lines are the moir\'{e} bands with $\xi=1$. }
\label{figs5}
\end{figure}

\vbox{}

\emph{Moir\'{e} bands calculated with full parameters model}---In Fig.~2  of the main text, the moir\'{e} bands are calculated with minimal model, where the influence of the remote hopping $\gamma_3$ and $\gamma_4$ are not considered. Here, we give some results based on full parameters model in Fig.~\ref{figs4},  in order to illustrate the more realistic situations. Fig.~\ref{figs4} (a-b) are for A+ABA, (c-d) are for AB+ABA and (e-f) are for ABA+ABA. The first (second) row in Fig.~\ref{figs4} is at $\theta=1.33^\circ$ ($1.05^\circ$).
First of all, due to the remote hopping,  the flat bands become dispersive but their bandwidth are still very small (red  solid lines). Meanwhile, the two  flat bands are separated in energy, \textit{e.g.}, see Fig.~\ref{figs4} (c,d).
Second, a twist angle dependent gap at the Dirac point of the linear bands is induced by $\gamma_3$ and $\gamma_4$ (blue solid lines). In Fig.~\ref{figs4}, the dashed lines are the moir\'{e} bands with $\xi=1$.

\vbox{}

\emph{Tunable valley Chern number}---In Fig.~3 of the main text, we show that, in ABA+ABA and BAB+ABA, a proper perpendicular electric field can give rise to four isolated nearly flat bands, which have nonzero valley Chern number.
The valley Chern number is calculated by the standard formula $C_n=\int_{mBZ}d^2\bm{k}\Omega_n(\bm{k})/2\pi$, where n is the band index. The Berry connection is
\renewcommand\theequation{S1}\begin{equation}
\begin{aligned}
     \Omega_n(\vec{k})=-2\sum_{n\neq n'}
\textrm{Im}[\frac{\langle u_n|\frac{\partial H}{\partial k_x}|u_{n'}\rangle\langle u_{n'}|\frac{\partial H}{\partial k_y}|u_{n}\rangle}{(E_{n'}-E_{n})^2}],
\end{aligned}
\end{equation}
where $|u_{n}\rangle$ is the moir\'e superlattice Bloch state, and  $E_n$ is the corresponding  eigenvalue.

Such topological flat bands can also be realized in AB+ABA and BA+ABA tFL-graphite, which is shown in Fig.~\ref{figs5}. Fig.~\ref{figs5} (a) is the moir\'{e} bands of AB+ABA under an electric field $V=20$ meV at $\theta=1.33^\circ$, where the valley Chern number of each band is denoted by the black number. That of BA+ABA is given in Fig.~\ref{figs5} (b) with $V=20$ meV and $\theta=1.33^\circ$. We see that, though AB+ABA and BA+ABA have similar band structure, but their valley Chern number are different.

The valley Chern number depends on the twist angle and the perpendicular electric field. As an example, we calculate the moir\'{e} bands and the valley Chern number of ABA+ABA with various potential difference [V=10, 20 meV] and different twist angle [$\theta=1.33^\circ$, $1.05^\circ$], see in Fig.~\ref{figs5} (c-f). Similar phenomenon has been reported in the twisted multilayer graphene with rhombohedral stacking, see Ref. [33-35, 40, 45] of the main text.

Note that in thicker tFL-graphite with  $N>3$, we can not get isolated moir\'{e} bands even if  a perpendicular electric field is applied , and thus there is no well defined valley Chern number for a single moir\'{e} band.

\section{III.Complementary results of the  thicker tFL-graphite with $ N>3$}
\begin{figure}
\centering
\includegraphics[width=14.5cm]{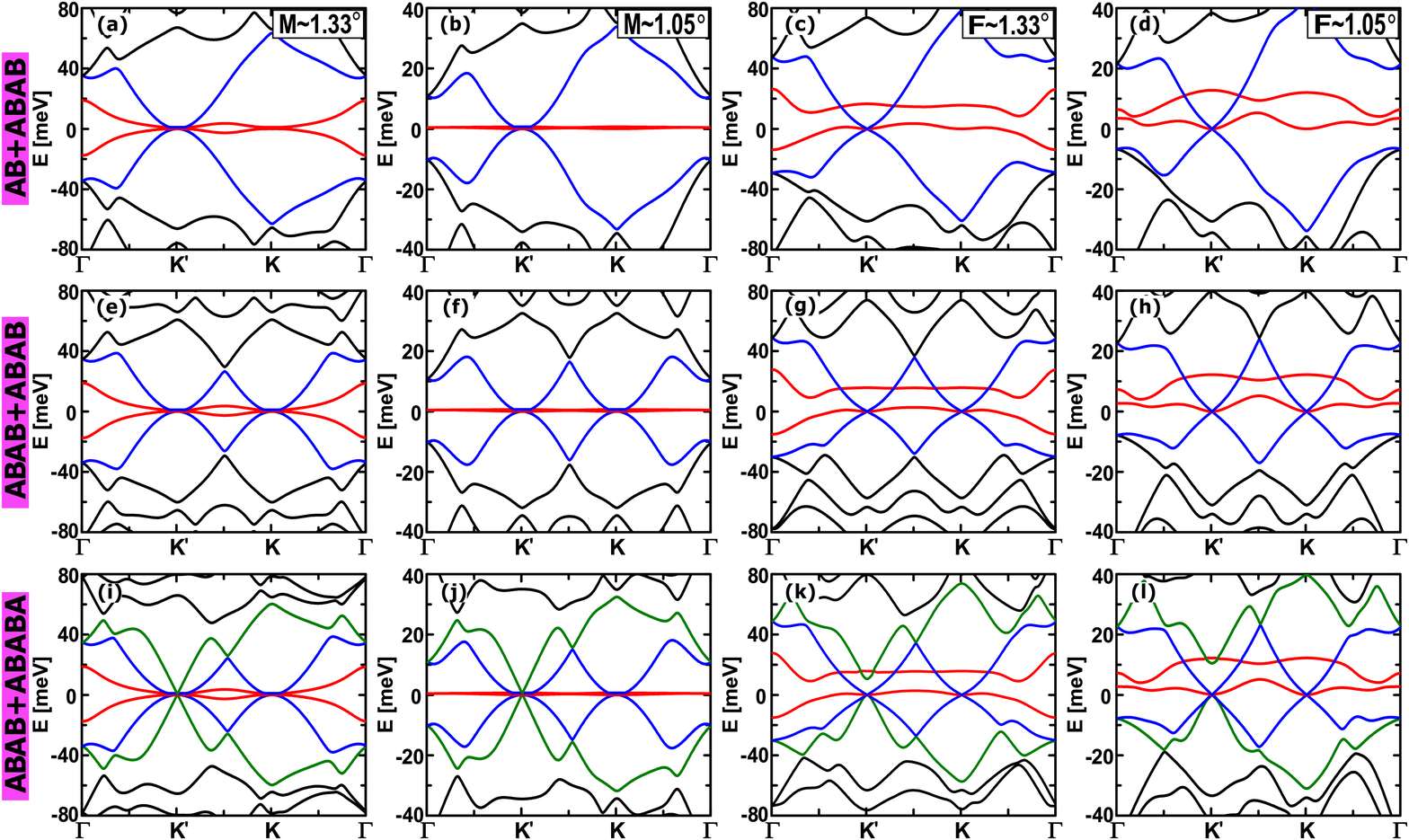}
\renewcommand\thefigure{S\arabic{figure}}
\caption{(a-d), (e-h) and (i-l) are the moir\'{e} bands of the AB+ABAB, ABAB+ABAB and ABAB+ABABA, respectively. (a, e, i) are calculated with the minimal model at $\theta=1.33^\circ$; (b, f, j) are calculated with minimal model at $\theta=1.05^\circ$; (c, g, k) are calculated with full parameter model at $\theta=1.33^\circ$; (d, h, l) are calculated with full parameter model at $\theta=1.05^\circ$. The parameters are the same as the Fig.~4 of the main text.}
\label{figs6}
\end{figure}

\emph{Moir\'{e} bands of several related tFL-graphites}---In the main text, we have studied the moir\'e bands of  thicker tFL-graphites with $N>3$. Several typical examples are given in Fig.4, \textit{i.e.},~BA+ABAB, BABA+ABAB, BABA+ABABA.
Note that there are several tFL-graphite configurations closely related to these examples, \textit{i.e.}, AB+ABAB, ABAB+ABAB, ABAB+ABABA, where the only difference is the relative orientation of the vdW layers.  For comparison, we plot the moir\'{e} bands of these related tFL-graphites in Fig.~\ref{figs6}.

\vbox{}

\begin{figure}
\centering
\includegraphics[width=10cm]{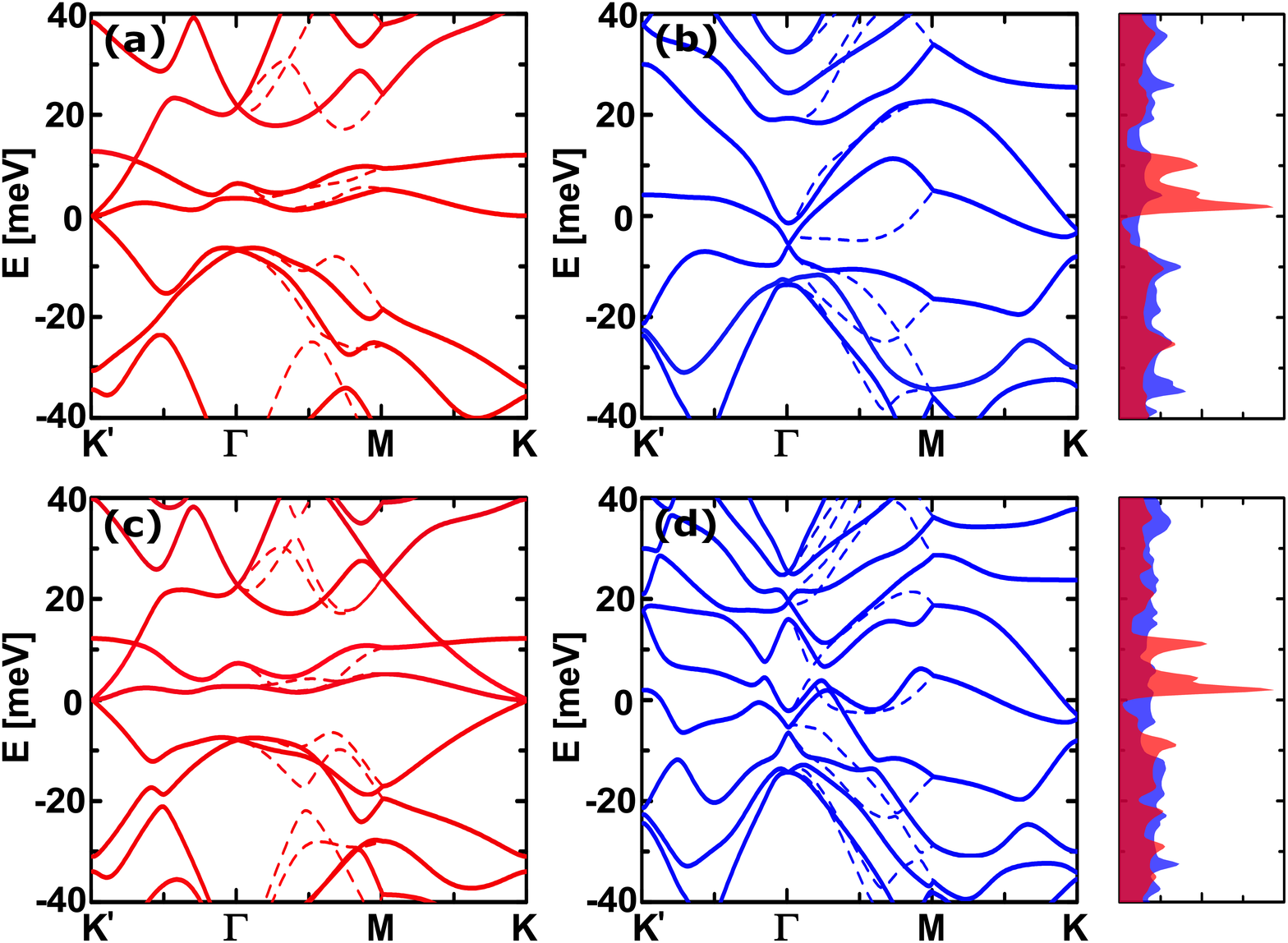}
\renewcommand\thefigure{S\arabic{figure}}
\caption{(a) is the moir\'e bands of the AB+ABAB  with zero electric field at $\theta=1.05^\circ$ (red), and  (b) is that  with $V=20$ meV (blue). (c) is the moir\'e bands of the ABAB+ABAB  with zero electric field at $\theta=1.05^\circ$ (red), and  (d) is that  with $V=20$ meV (blue). The right column is the corresponding DOS, where red is for zero electric field and blue is for finite electric field. Full parameter model is used: $\gamma_0/\gamma_1/\gamma_3/\gamma_4=2610/360/283/138$ meV.}
\label{figs7}
\end{figure}

\emph{Electric field influence on the moir\'{e} bands}---In Fig.~3 of the mian text, we show that a perpendicular electric field can significantly change the moir\'{e} bands in the cases of $N=3$.
Actually, a perpendicular electric field also can dramatically modify the moir\'{e} bands when $N>3$. Two examples, \textit{i.e.}, AB+ABAB, ABAB+ABAB, are given in Fig.~\ref{figs7}. As a comparison, we first plot the moir\'{e} bands of AB+ABAB and ABAB+ABAB with zero electric field  at $\theta=1.05^\circ$ in Fig.~\ref{figs7} (a) and (c), respectively (red lines). Then, an electric field with $V=20$ meV is applied, and the corresponding moir\'{e} bands are given in Fig.~\ref{figs7} (b) and (d), see the blue lines. We see that the band shape, as well as the DOS, is significantly changed by the applied electric field.


\end{appendix}

\end{document}